\pdfoutput=1
\documentclass[a4paper,showkeys,amsmath,amssymb,superscriptaddress,pra,twocolumn,floatfix,showpacs]{revtex4}
\usepackage[utf8]{inputenc}
\usepackage{amsmath,amssymb}
\usepackage{multirow}
\usepackage{graphicx}
\usepackage{subfig}
\usepackage{xcolor}
\usepackage{bbm}
\usepackage{hyperref}

\newcommand{\ket}[1]{| #1 \rangle}
\newcommand{\bra}[1]{\langle #1 |}
\newcommand{\ketbra}[2]{| #1 \rangle \langle #2 |}
\newcommand{\braket}[2]{\bra{#2}{#1}\rangle}
\newcommand{\proj}[1]{| #1 \rangle \langle #1 |}

\newcommand{\tr}{\mathrm{Tr}}

\newcommand{\Tr}{\mathrm{Tr}}
\newcommand{\ee}{\mathrm{e}}
\newcommand{\ii}{\mathrm{i}}
\newcommand{\dd}{\mathrm{d}}
\newcommand{\1}{\mathbbm{1}}

\newcommand{\R}{\mathbb R}
\newcommand{\C}{\mathbb C}

\newcommand{\MP}{\mathcal{MP}}
\newcommand{\SMP}{\mathcal{SMP}}

\newtheorem{lemma}{Lemma}
\newtheorem{proposition}{Proposition}

\begin{document}
\author{Zbigniew Pucha{\l}a}
\affiliation{Institute of Theoretical and Applied Informatics, Polish Academy
of Sciences, ulica Ba{\l}tycka 5, 44-100 Gliwice, Poland}
\affiliation{Institute of Physics, Jagiellonian University, ulica Stanis\l{}awa 
\L{}ojasiewicza 11, 30-348 Krak\'o{}w, Poland}

\author{{\L}ukasz Pawela\footnote{Corresponding author, E-mail:
lpawela@iitis.pl}} % % \email{lpawela@iitis.pl}
\affiliation{Institute of Theoretical and Applied Informatics, Polish Academy
of Sciences, ulica Ba{\l}tycka 5, 44-100 Gliwice, Poland}

\author{Karol {\.Z}yczkowski}
\affiliation{Institute of Physics, Jagiellonian University, ulica
Stanis\l{}awa \L{}ojasiewicza 11, 30-348 Krak\'o{}w, Poland}
\affiliation{Center for Theoretical Physics, Polish Academy of Sciences, Aleja
Lotnik{\'o}w 32/46, 02-668 Warsaw, Poland}

\title{Distinguishability of generic quantum states}
\date{May 12, 2016}

\begin{abstract}
Properties of random mixed states of dimension $N$ distributed uniformly with
respect to the Hilbert-Schmidt measure are investigated. We show that for large
$N$, due to the concentration of measure, the trace distance between two random
states tends to a fixed  number ${\tilde D}=1/4+1/\pi$, which yields the
Helstrom bound on their distinguishability. To arrive at this result we apply
free random calculus and derive the symmetrized Marchenko--Pastur distribution,
which is shown to describe numerical data for the model of coupled quantum
kicked tops. Asymptotic value for the root fidelity between two random states,
$\sqrt{F}=3/4$, can serve as a universal reference value for further
theoretical and experimental studies.
Analogous results for quantum relative entropy and Chernoff quantity provide
other bounds on the distinguishablity of both states in a multiple measurement
setup due to the quantum Sanov theorem.
We study also mean entropy of coherence of random pure and mixed states 
and entanglement of a generic mixed state of a bi--partite system.
\end{abstract}

\pacs{02.50.Cw, 03.67.-a, 05.45.Mt}

\keywords{quantum distinguishability, random states, mean fidelity,
           trace distance, concentration of measure, Helstrom bound}

\maketitle
%%%%%%%%%%%%%%%%%%%%%%%%%%%%%%%%%%%%%%%%%%%%%%%%%%%%%%%%%%%%%%%%%%%%%%%%%%%%%%%%
\section{Introduction}
%%%%%%%%%%%%%%%%%%%%%%%%%%%%%%%%%%%%%%%%%%%%%%%%%%%%%%%%%%%%%%%%%%%%%%%%%%%%%%%%
Processing of quantum information takes place not in a Hilbert space but in a
laboratory \cite{Pe95}. However, the standard formalism of density operators
acting on a finite dimensional Hilbert space ${\cal H}_N$ proves to be
extremely efficient in describing physical experiments. It is therefore
important to investigate properties of the set $\Omega_N$ of density operators
of a given size $N$, as it forms a scene for which screenplays for quantum
pieces are written.

In the one-qubit case the set $\Omega_2$ forms the familiar Bloch ball, but the
geometry of the set $\Omega_N \subset {\mathbbm R}^{N^2-1}$ for $N> 3$ becomes
much more complex  \cite{BZ06}. Interestingly, for large $N$ certain
calculations become easier due to the measure concentration phenomenon: a
slowly varying function of a random state typically takes values close to the
mean value \cite{Le01,HLW06}. Extending the approach of \cite{Mo07,AL15} we
focus our attention on the distinguishability between generic quantum states.

Two states $\rho$ and $\sigma$, can be distinguished by a suitable experiment
with probability one if they are orthogonal, \emph{i.e.} $\tr \rho \sigma = 0$.
The celebrated result of Helstrom \cite{He69} provides a bound for the
probability $p$ of the correct distinction between arbitrary two  quantum
states in a single experiment,
\begin{equation}
p \; \leq \; \frac12 \bigl[ 1 + D_{\rm Tr}(\rho,\sigma) \bigr],
\end{equation}
where the definition of the \emph{trace distance} reads
\begin{equation}
D_{\rm Tr}(\rho,\sigma)= \frac12 {\rm Tr}|\rho - \sigma | .
\label{traced}
\end{equation}
For two orthogonal states this distance is maximal, $D_{\rm
Tr}= 1$,  so $p=1$ as expected. The above bound can be achieved by a projective
measurement onto positive and negative part of the \emph{Helstrom matrix},
$\Gamma = \rho - \sigma$.

In the case of measurement on $m$ copies of both states the probability of an
error decreases exponentially with the number of copies. To characterize the
optimal distinguishability rate one can apply the quantum Stein
lemma~\cite{HP91}, the Sanov theorem \cite{BD+08} and the relative entropy
between both states, or the quantum Chernoff bound \cite{Cas08}.

The aim of this work is to analyze properties of generic quantum states in high
dimensions and their distinguishability. For two random states distributed
according to the flat, Lebesgue measure in the set $\Omega_N$ of quantum states
of dimension $N\gg 1$, we derive the limiting values of fidelity, trace
distance, relative entropy and Chernoff quantity, which allow us to obtain
universal bounds for their distingushability. Furtheremore, we study average
coherence of a random state with respect to a given basis and average
entanglement of random mixed state of a bipartite quantum system.

%%%%%%%%%%%%%%%%%%%%%%%%%%%%%%%%%%%%%%%%%%%%%%%%%%%%%%%%%%%%%%%%%%%%%%%%%%%%%%%%
\section{Random quantum states and their level density}
%%%%%%%%%%%%%%%%%%%%%%%%%%%%%%%%%%%%%%%%%%%%%%%%%%%%%%%%%%%%%%%%%%%%%%%%%%%%%%%%
Consider a Haar random unitary matrix $U$ of dimension $NK$ acting on a fixed
bipartite state $|\phi\rangle \in {\cal H}_N \otimes {\cal H}_K$. Then the state
$|\psi\rangle =U|\phi\rangle$ is distributed according to the Haar measure on
the space of pure states, while its partial trace $\sigma={\rm Tr}_K
\ketbra{\psi}{\psi}$ is distributed in the set $\Omega_N$ according to
the induced measure $\nu_K$ \cite{ZS01}. To be specific, the integral with
respect to the induced measure $\nu_K$ on a set of quantum mixed states is
equal to the integral of the partial trace over a group of unitary matrices of
size $N \times K$ equipped with the normalized Haar measure,
\begin{equation}
\int_{\Omega_N} f(\rho) \dd \nu_K(\rho)
=
\int f\left(\tr_K (U \ketbra{0 0 }{0 0} U^{\dagger}) \right) \dd \mu_H(U).
\end{equation}
In the special case of $K = N$ we get the flat Hilbert Schmidt measure,
$\nu_{HS}=\nu_N$. In particular, setting $K=N=2$ one generates the flat
Lebesgue measure on the set $\Omega_2$ of one qubit mixed states equivalent to
the Bloch ball. Alternatively, such random states can be constructed as
$\rho=GG^{\dagger}/{\rm Tr} GG^{\dagger}$, where $G$ stands for a random
rectangular matrix of dimension $N \times K$ with independent complex normal
entries. In the case where $K=N$ large, the flat measure in the set $\Omega_N$
leads asymptotically to the Marchenko--Pastur distribution
$\MP(x)$~\cite{MP67}. In the general case this distribution depends on the
rectangularity parameter $c=K/N$. For $c \geq 1$ the density is continuous,
\begin{equation}
\MP_c (x) = \frac{1}{2 \pi  x}
\sqrt{\left[x-(1-\sqrt{c})^2\right] \left[(1+\sqrt{c})^2-x \right]},
\label{eq:fc}
\end{equation}
while for $c < 1$ the measure also contains a singular component and reads $(1
- c)\delta_0 + \MP_c(x)$. 

Here $x = K\lambda$ is a rescaled eigenvalue of
$\rho$. In the case where $c=1$ the $\MP$ measure $\mu_{\MP}$ has a continuous
density, $\MP(x)=(\sqrt{4/x-1})/2\pi$, supported on $[0,4]$.

Consider a random quantum state $\rho$ generated according to the flat measure
$\nu_{\rm HS}$ in $\Omega_N$ with eigenvalues $\lambda_i$ described by the
measure $\mu_N^{(\rho)} = \frac1N \sum_i \delta_{N \lambda_i(\rho)}$. For a
large $N$ with probability one the measure $\mu_N$ converges weakly to
$\mu_{\MP}$. Thus, for any bounded and continuous matrix function $g$ of a
random state $\rho$, the trace of $g(\rho)$ converges almost surely ($a.s.$) to
the mean value,
\begin{equation}
\tr( g( \rho )) = N \!\int \! g\left(\frac{t}{N}\right) \dd \mu_N^{(\rho)}(t)
\xrightarrow[{N\to \infty}]{a.s.}
\int \tilde{g}(t) \dd \mu_{\MP}(t),
\label{integra}
\end{equation}
where $\tilde{g}(t) = \lim_{N\to\infty}Ng(\frac tN)$.
%%%%%%%%%%%%%%%%%%%%%%%%%%%%%%%%%%%%%%%%%%%%%%%%%%%%%%%%%%%%%%%%%%%%%%%%%%%%%%%%
\section{Trace distance and a single shot experiment}
%%%%%%%%%%%%%%%%%%%%%%%%%%%%%%%%%%%%%%%%%%%%%%%%%%%%%%%%%%%%%%%%%%%%%%%%%%%%%%%%
To show a simple application of Eq. (\ref{integra}) consider the trace distance
of a random state $\rho$ to the maximally mixed state $\rho_*=\1 /N$. For a
large dimension $N$ the value
\begin{equation}
\tr \left|\rho - \rho_*\right| = \int |t - 1| \dd \mu_N^{(\rho)}(t),
\end{equation}
converges to the integral over the $\MP$ measure, so that the trace distance
behaves almost surely as
\begin{equation}
D_{\rm Tr}(\rho,\rho_*)
\xrightarrow[{N\to \infty}]{a.s.}
 \frac12\int |t - 1| \dd \mu_{\MP}(t)
= \frac{3\sqrt{3}}{4\pi} \simeq 0.4135.
\label{avesingle}
\end{equation}

We wish to evaluate the trace distance between two random quantum states $\rho,
\sigma$ sampled from the set $\Omega_N$ according to an induced measure
$\nu_K$. To this end we need to derive the spectral density of the Helstrom
matrix $\Gamma=\rho-\sigma$, for which we use tools of free probability. In
particular, we are going to apply the free additive convolution of two
distributions \cite{voiculescu1986addition}, written $ P_1 \boxplus P_2$, which
describes the asymptotic spectrum of the sum of two random matrices $X_1$ and
$X_2$, each described by the densities $P_1$ and $P_2$, under the assumption
that both variables are independent and invariant with respect to unitary
transformations, $X_i \to UX_i U^{\dagger}$. As both random states $\rho$ and
$\sigma$ are independent and free, the limiting spectral density of their
difference is given by the free additive convolution %
% ~\cite{voiculescu1986addition}
of the Marchenko--Pastur distribution and its dilation $\cal D$ by factor minus
one, $\SMP =  \MP \boxplus {\cal D}_{-1}({\MP})$. In order to derive $\SMP$, it
is convenient to use the $R$ transform of the $\MP$ distribution, given by:
\begin{equation}
R(G(z)) + \frac{1}{G(z)} = z,
\end{equation}
where $G(z)$ is the Cauchy transform of the distribution \cite{nica1998commutators}.
 Hence, we get that
the $R$-transform of the MP distribution reads 
$R_{\MP}(z) =  c /(1 - z)$ so the
$R$-transform of the symmetrized distribution is given by the sum,
\begin{equation}
R_{\SMP}(z) =  R_{\MP}(z) - R_{\MP}( - z ) = \frac{2 c   z}{1 - z^2}.
\end{equation}
Inverting this transform we obtain the desired \emph{Symmetrized
Marchenko-Pastur} (SMP) distribution:
\begin{equation}
\label{eq:sym-mp-dist}
\SMP_{c}(y) = \frac{
-1 -4 (c -1)  c -3 y^2+[Y_c(y)]^{2/3}}
{2 \sqrt{3} \pi  y [Y_c(y)]^{1/3}},
\end{equation}
where 
\begin{equation}
\begin{split}
Y_c(y) = & (2 c -1)^3+9 (c +1) y^2+3 \sqrt{3} y \\
& \times \sqrt{(2  c -1)^3 - y^4+[2-( c -10) c ] y^2} .
\end{split} 
\end{equation}
This measure is supported between $y_\pm = \pm
\frac{1}{\sqrt{2}}\sqrt{-c^2+10 c +(c +4)^{3/2} \sqrt{c}+2}$. If $c<1$ the
measure has a singularity at zero.

\begin{figure}[!h]
\subfloat{\centering\includegraphics[width=0.23\textwidth]{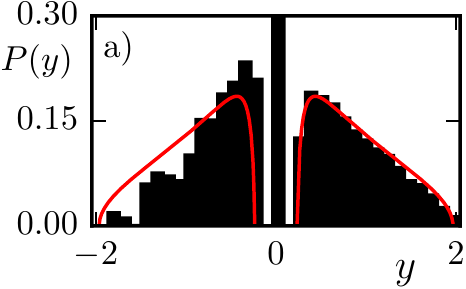}}
\subfloat{\centering\includegraphics[width=0.23\textwidth]{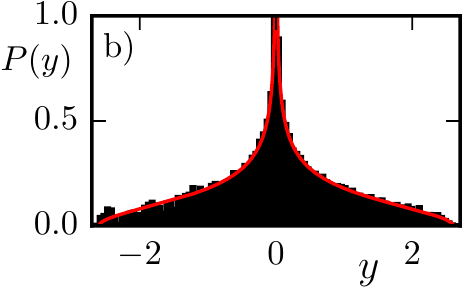}}\\[0.1cm]
\subfloat{\centering\includegraphics[width=0.23\textwidth]{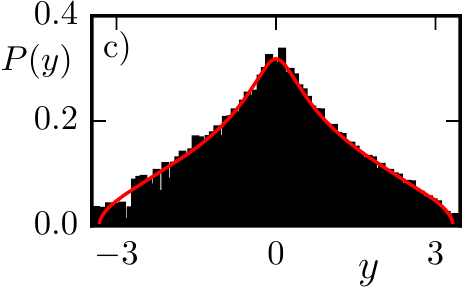}}
\subfloat{\centering\includegraphics[width=0.23\textwidth]{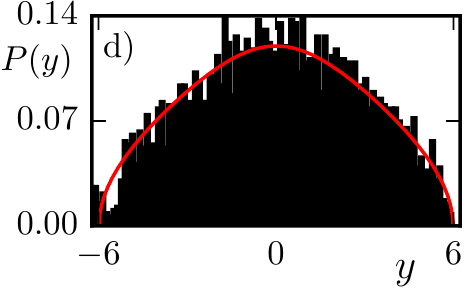}}
\caption{Symmetrized Marchenko--Pastur distribution \eqref{eq:sym-mp-dist} for
$c=0.2 \;(a), 0.5 \;(b), 1.0 \;(c), 4.0\;(d)$ denoted by the solid (red) line.
Histograms represent % Monte Carlo simulation
numerical results % obtained
for the model of coupled
kicked tops (\ref{2tops}, \ref{eqn:traced-tops}) with $N=100$ cumulated
out of $100$ realizations of the system.}
\label{fig1}
\end{figure}

In the special case of $c = 1$, corresponding to the Hilbert-Schmidt
measure $\nu_{\rm HS}$ in  $\Omega_N$, we obtain
\begin{equation}
\!\!\!\!
\SMP(y) \!= \!\!\frac
{-1 -3 y^2+\left(1 + 3 y \left(\sqrt{3+ 33 y^2 -3 y^4}+6 y\right)\right)^{2/3}}
{2 \sqrt{3} \pi  y \left(1 + 3 y \left(\sqrt{3+ 33 y^2 -3 y^4}+6  
y\right)\right)^{1/3}}.
\end{equation}
This measure was earlier identified as a free commutator between two
semicircular distributions~\cite{nica1998commutators} and 
appeared in the literature 
under the name of the tetilla law~\cite{deya2012convergence}.

As a simple application of the $\SMP$ distribution  we will sow that the 
Hilbert-Schmidt distance tends to zero as $N$ tends to infinity.

To obtain the required result consider two independent random states $\rho$ and 
$\sigma$, distributed according to the Hilbert-Schmidt measure and rewrite 
Eq.~\eqref{integra} as:
\begin{equation}
\tr( g( \rho -\sigma )) = N \!\int \! g\left(\frac{t}{N}\right) \dd 
\mu_N^{(\rho-\sigma)}(t)
\end{equation}
From this follows:
\begin{equation}
\| \rho - \sigma \|_2^2 = 
\tr(\rho - \sigma)^2 = \frac1N 
\!\int \! t^2 \dd \mu_N^{(\rho-\sigma)}(t).
\end{equation}
Now taking the limit $N\to\infty$, the integral with probability one has a
finite value $\int \! t^2 
\dd \mu_\SMP(t)=2$, hence due to the prefactor $N^{-1}$
we get asymptotically the desired result,
\begin{equation}
\label{HS3}
 D_{\rm HS}(\rho,\sigma) =\| \rho - \sigma \|_2 \simeq \sqrt{\frac{2}{N}}
\xrightarrow[{\tiny{N\to \infty}}]{a.s.} 0 \; .
\end{equation}
An analogous result implying that the average HS distance $D_{\rm
HS}(\rho,\rho_*)$ between  a random state $\rho$ and the maximally mixed state
$\rho_*$ for large dimension $N$ tends to zero was earlier obtained in
\cite{HLW06}. 
Although this distance is not monotone \cite{BZ06} it admits a
direct operational interpretation \cite{LKB03}.

The asymptotic value of ${\rm Tr}|\rho-\sigma|={\rm Tr}|\Gamma|$ will be almost
surely given by an integral over the $\SMP$ distribution, which yields the
generic trace distance between two random  states,
\begin{equation}
D_{\rm Tr}(\rho,\sigma)
\xrightarrow[{N\to \infty}]{a.s.}
{\tilde D}=\frac12\int |y| \dd \mu_{\SMP}(y)
= \frac{1}{4} + \frac{1}{\pi} \simeq 0.5683.
\label{aveboth}
\end{equation}

Formulae  (\ref{avesingle}) and (\ref{aveboth}) 
combined with the Helstrom theorem
imply one of the key results of this work.

\begin{proposition}.
Consider two independent random states $\rho$ and
$\sigma$ drawn according to the HS measure in the space of states of dimension
$N$. For large $N$ the probability $p_2$ of correct distinction between both
random states in a single measurement is bounded by
$
p_2 \leq  \frac12 + \frac12 \left(\frac14 + \frac{1}{\pi}\right) \simeq 
0.7842,
$
while the probability $p_1$ of distinguishing $\sigma$ and 
the maximally mixed state $\rho_*=\1/N$ is bounded by
$
p_1 \leq  \frac12 + \frac12 \frac{3\sqrt{3}}{4\pi} \simeq 0.7067.
$
\end{proposition}

\begin{figure}[!h]
{%\centering
\includegraphics{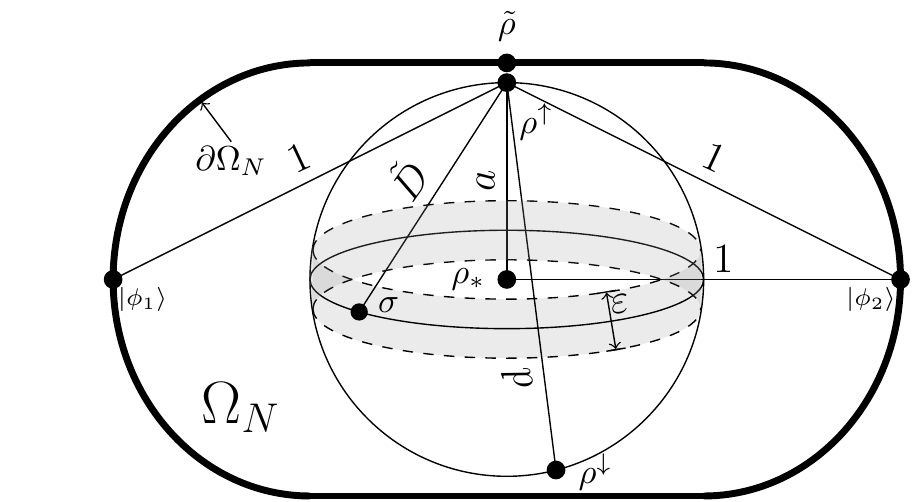}}
\caption{Sketch of the set $\Omega_N$ of mixed states for large dimension: the
measure is concentrated along the unitary orbit of a generic state
$\rho=\rho^{\uparrow}$. Its trace distance to the center $\rho_*$ of the set is
$a \simeq 0.41$ and to another typical state
 $\sigma$ is ${\tilde D} \simeq 0.57$.}
\label{fig2}
\end{figure}

The above results improve our understanding \cite{BZ06} of the structure of the
set $\Omega_N$ of mixed states of a large dimension $N$. The HS measure is
concentrated in an $\varepsilon$ neighborhood of the unitary orbit, $U \rho
U^{\dagger}$, where $U$ is unitary and $\rho$ is a random mixed quantum state
with spectrum  distributed according to $\mu_{\MP}$. The width of the orbit
decreases as $\frac1N$ -- this will be shown in the next section -- while its
diameter is given by the distance between two diagonal matrices with opposite
order of the eigenvalues \cite{MMPZ08}, $d=D_{\rm Tr} (p^{\uparrow},
p^{\downarrow})= \int_0^4 x  \, {\rm sign}(x-M)d \mu_{\MP}(x)\simeq 0.7875$,
where $M$ denotes the median,  $\int_0^M d \mu_{\MP}=1/2$. A generic state
$\rho$ is located at the distance $a= \frac{3\sqrt{3}}{4\pi}$ from the center
$\rho_*=\frac{\1}{N}$, its distance to the closest boundary state $\tilde \rho$
tends to zero, while the distances to the closest and to the most distant pure
state, $\ket{\phi_1}$ and $\ket{\phi_2}$, tend to unity -- see Fig. \ref{fig2}.

To study the case of a large environment, $K \gg N$, we take the limit $c \to
\infty$ of Eq.~\eqref{eq:sym-mp-dist}. This distribution, rescaled by a factor
$1/\sqrt{2c}$ tends to the Wigner semi-circle law $C(y)$ supported on $[-2,2]$,
as its $R$-transform converges to $R_W(z)=z$. The average $|y|$ over the
semicircle is finite in analogy to ~\eqref{aveboth}, but due to the scaling
factor $c^{-1/2}$ the typical trace distance with respect to the limiting
measure $\nu_\infty$ tends to $0$ -- see Appendix \ref{appenB}.

Note that the trace distance asymptotically tends to a fixed
value~\eqref{aveboth}, while the average HS distance tends to zero. This is a
consequence of the known fact that if the taxi distance between two points in
${\mathbbm R}^{n+1}$ is fixed, their Euclidean distance is typically much
smaller, since shortcuts across city blocks are allowed. More formally, one can
show that a random point from the unit sphere $S^n$ with respect to the taxi
metric $L_1$ is located at the Euclidean distance $L_2\sim 1/\sqrt{n}$ form the
center of the sphere. A similar relation between the trace and the
Hilbert-Schmidt distance holds in the space of quantum states and explains
result (\ref{HS3}).

To visualize this, we present here, for comparison, the average distances
between two random points in a unit ball in $\mathbb{R}^n$ distributed
according to the Euclidean measure. The distances with respect to the norms
$L_1$, $L_2$ and $L_{\infty}$ are provided in Tab.~\ref{tab:ball-dists}. Note
that the definition of the trace distance contains the prefactor $1/2$, so the
$L_1$ distance has to be compared with $2 D_{\rm Tr}$. For large dimensions the
full mass of the $n$--ball is concentrated close to its surface, so the $L_1$
distance between two typical points diverges, while in the case of the set of
mixed states, the layer of generic states remains distant from the subset of
pure states and the average asymptotic distance becomes finite and
deterministic.

\begin{table}[!h]
\begin{tabular}{|c|c|c|c|}
\hline
$n$ & $L_1$ & $L_2$ & $L_\infty$  \\ \hline
1 & $\frac23$ & $\frac23$ & $\frac23$ \\ \hline
2 & $\frac{647+120\pi}{90\pi^2} \simeq 1.1528$ & $\frac{128}{45\pi} \simeq 
0.9054$ & 0.8151 \\ \hline
3 & $\frac{55\pi}{112} \simeq 1.15428$ & $\frac{36}{35} \simeq 1.0286$ & 
0.8549\\ 
\hline
$\dots$ & $\dots$ & $\dots$ & $\dots$ \\ \hline
$\infty$ & $\infty$ & $\sqrt{2}$ & 0 \\ \hline
\end{tabular}
\caption{Average distances of points in a unit ball of dimension $n$
with respect  to the norms $L_1$, $L_2$ and $L_\infty$.}
\label{tab:ball-dists}
\end{table}
%%%%%%%%%%%%%%%%%%%%%%%%%%%%%%%%%%%%%%%%%%%%%%%%%%%%%%%%%%%%%%%%%%%%%%%%%%%%%%%%
\section{Application of Levy's lemma}
%%%%%%%%%%%%%%%%%%%%%%%%%%%%%%%%%%%%%%%%%%%%%%%%%%%%%%%%%%%%%%%%%%%%%%%%%%%%%%%%
In this section we present the Levy's lemma, used to strengthen our results and
to characterize the convergence rate of the trace distance between two random
states to the mean value $\tilde{D}$. Note that the key results from the
previous section, showing that various distances between two generic states
converge to their deterministic values, are independent of the Levy's Lemma.

\begin{lemma}[Levy's lemma]\!.
Let $f: S^{m-1} \to \mathbb{R}$ be a Lipschitz-continuous function with
Lipschitz constant $\eta$, i.e.
\begin{equation}
\Big| f(x) - f(y) \Big| \leq \eta \|x - y\|.
\end{equation}
For a random element of the sphere $x \in S^{m-1}$ distributed uniformly, for 
all $\varepsilon >0$ we get
\begin{equation}
P(|f(x) - \mathbb{E} f(x)| \geq \varepsilon) \leq 2 \exp\left( - \frac{m
\varepsilon^2}{9 \pi^3 \eta^2}\right).
\end{equation}
\end{lemma}

Following \cite{HLW06} we are going to use the  Levy's lemma for a set of
bipartite pure states of a high dimension. Consider two states
$\ket{\psi},\ket{\phi} \in {\cal H}_N \otimes {\cal H}_N$ and the function
\begin{equation}
f(\ket{\psi},\ket{\phi}) = D_{\tr}(\rho,\sigma),
\end{equation}
where $\rho=\tr_B\proj{\psi}$ and $\sigma=\tr_B \proj{\phi}$ represent reduced
states. For a fixed state $\ket{\phi}$ the above function has the Lipschitz
property with the Lipschitz constant equal to unity. To show this we write
\begin{equation}
\begin{split}
&
\Big|f(\ket{\psi},\ket{\phi}) - f(\ket{\xi},\ket{\phi}) \Big| = 
\Big| \frac12 \|\rho - \sigma \|_1 - \frac12 \|\rho' - \sigma \|_1 \Big|
\\
& \leq
\frac12 \|\rho - \rho' \|_1,
\end{split}
\end{equation}
where $\rho' =  \tr_B\proj{\xi}$.
Next we rely on the inequality of Fuchs and van de Graaf 
\cite{BZ06,FG99}, 
\begin{equation}
D_{\tr}(\rho,\rho') \leq  \sqrt{1 - F(\rho,\rho')},
\end{equation}
where the fidelity $F$ between two states reads  
\begin{equation}
F(\rho, \sigma) =  ({\rm Tr} |\rho^{1/2}\sigma^{1/2}|)^2 .
\label{fidel}
\end{equation}
%in the main body of the paper.
Using an elementary inequality $\sqrt{1-x^2} \leq
\sqrt{2-2x}$ for $x \in [0,1]$, we write
\begin{equation}
D_{\tr}(\rho,\rho') \leq  \sqrt{2 - 2 \sqrt{F(\rho,\rho')}}
\leq  \| \, \ket{\psi} - \ket{\xi} \, \|.
\end{equation}
The last inequality follows from the Uhlmann theorem \cite{BZ06} relating
fidelity to the maximal overlap of purifications. This consideration gives us,
that for a fixed $\ket{\phi}$ the function $f(\ket{\psi},\ket{\phi})$ is
$1$--Lipschitz.

Using the Levy's lemma, for a fixed state $\ket{\phi}$ and $\ket{\psi}$
distributed uniformly on a set of pure states in  ${\cal H}_N \otimes {\cal
H}_N$ we obtain
\begin{equation}
\begin{split}
&P\left( \Big|f(\ket{\psi},\ket{\phi}) - \mathbb{E} f(\ket{\psi},\ket{\phi}) 
\Big| > \varepsilon\right)\\
&=P\left( \Big| 
D_{\tr}(\rho,\sigma)  - \mathbb{E} D_{\tr}(\rho,\sigma) \Big|  > 
\varepsilon\right) \\
&\leq 2 \exp\left( - \frac{N^2 \varepsilon^2}{9 \pi^3 }\right).
\end{split}
\end{equation}
Let us now consider a sequence of fixed states $\{\ket{\phi_N}\}_N$, such that
the empirical distributions of eigenvalues of the reduced states $\sigma_N$ 
tend to the Marchenko-Pastur law. For a random state $\ket{\psi} \in {\cal
H}_N \otimes {\cal H}_N$ with $\rho = \tr_B \proj{\psi}$ we define an
expectation value with respect to the distribution of $\ket{\psi}$,
\begin{equation}
\Delta_N = 
\mathbb{E} f(\ket{\psi},\ket{\phi_N}) = 
\mathbb{E} D_{\tr}(\rho,\sigma_{N}).
\end{equation}
Unitary invariance of the trace distance and unitary invariance of the
distribution of $\rho$ implies that $\Delta_N \to \tilde{D} = \frac14 + 
\frac1\pi$.
The triangle inequality gives us the following set of inclusion relations
$
\left\{\rho: \Big| D_{\tr}(\rho,\sigma_N) - \Delta_N \Big| + \Big|\Delta_N - 
\tilde{D}\Big| < \varepsilon \right\} 
\subset 
\left\{\rho: \Big| D_{\tr}(\rho,\sigma_N) - \tilde{D}\Big| < \varepsilon 
\right\}.
$
For dimension $N$ so large that $|\Delta_N-\tilde{D}| < \varepsilon$
this implies the following bounds, 
\begin{equation}
\begin{split}
&P\left( \Big| D_{\tr}(\rho,\sigma_{N})) - \tilde{D} \Big|  < 
\varepsilon\right) \\
&\geq 
P\left( \Big| D_{\tr}(\rho,\sigma_{N})) - \Delta_N \Big| + 
|\Delta_N-\tilde{D}|  < \varepsilon\right) \\
&=
P\left( \Big| D_{\tr}(\rho,\sigma_{N})) - \Delta_N \Big|   < \varepsilon - 
|\Delta_N-\tilde{D}|\right) \\
&\geq 1 - 2 \exp\left( - \frac{N^2 (\varepsilon-|\Delta_N-\tilde{D}|)^2}{9 
\pi^3 }\right).
\end{split}
\end{equation}
Since $|\Delta_N-\tilde{D}| = o(1)$ we arrive at the following result.

\begin{proposition}.
Consider two independent random states $\rho$ and $\sigma$ distributed
according to the Hilbert-Schmidt measure  in the set $\Omega_N$ of quantum
states of dimension $N$. For sufficiently large $N$, when $|\mathbb{E} 
D_{\Tr}(\rho,
\sigma)-\tilde{D}|<\frac{\varepsilon}{2}$, their trace distance is close to the
number $\tilde D= \frac{1}{4} + \frac{1}{\pi}$, as the deviations become
exponentially rare,
\begin{equation}\label{eqn:epsilon-strip}
P\left( \Big| D_{\tr}(\rho,\sigma)) - \tilde{D} \Big|  >  \varepsilon\right) 
\leq 2 \exp\left( - \frac{N^2 (\varepsilon/2)^2}{9 \pi^3 }\right).
\end{equation}
\end{proposition}

The formal statement above can be interpreted that the orbit of generic quantum
states containing the entire mass of the set $\Omega_N$, represented in
Fig.~\ref{fig2} by an $\varepsilon$-strip, for large $N$ becomes
infinitesimally narrow as $\varepsilon \sim 1/N$. Since the exponent in
Eq.~\eqref{eqn:epsilon-strip} behaves as $N^2 \varepsilon^2$, the width
$\varepsilon$ of the strip decays with the dimension faster than $N^{\delta-1}$
  %$\frac{1}{N^{1-\delta}}$
for any $\delta>0$.
%%%%%%%%%%%%%%%%%%%%%%%%%%%%%%%%%%%%%%%%%%%%%%%%%%%%%%%%%%%%%%%%%%%%%%%%%%%%%%%%
\section{Several measurements, relative entropy and Chernoff bound}
%%%%%%%%%%%%%%%%%%%%%%%%%%%%%%%%%%%%%%%%%%%%%%%%%%%%%%%%%%%%%%%%%%%%%%%%%%%%%%%%
Consider the Chernoff function \cite{Cas08,ANSV08} of two  random states,
$Q_s(\rho, \sigma)= \tr \rho^s \sigma^{1-s}$ and the Chernoff information
$Q:=\min_{s\in[0, 1]}Q_s$. Based on the discussion in
Appendix~\ref{supp:sec:functions} the Chernoff function can be asymptotically
represented as a product
of two integrals over the Marchenko-Pastur measure,
\begin{equation}
\begin{split}
Q_s & \xrightarrow[{\tiny{N\to \infty}}]{a.s.}
\int t^s \dd \mu_{\MP}(t)  \times \int t^{1-s} \dd \mu'_{\MP}(t)  \\
&=
\frac{4 \Gamma \left(\frac{3}{2}-s\right) \Gamma 
\left(s+\frac{1}{2}\right)}{\pi  \Gamma (3-s) \Gamma (s+2)}.
\end{split}
\end{equation}

The above expression, involving the Gamma function
$\Gamma(x)$, takes its minimal value at 
$s = \frac12$, so that the generic value of the quantum Chernoff information
$Q$ reads
\begin{equation}
Q=\min_s \Bigl( \lim_{N\to\infty} Q_s \Bigr)
= \left(\frac{8}{3 \pi}\right)^2
\simeq 0.7205.
\label{chernoff}
\end{equation}

The Kullback-Leibler relative entropy reads $S(\rho \| \sigma) = {\rm Tr} \rho
\left(\log \rho - \log \sigma \right).$ 
Applying again the reasoning presented 
in Appendix~\ref{supp:sec:functions} % in SM
we get:
\begin{equation}
\begin{split}
S(\rho \| \sigma) \xrightarrow[{\tiny{N\to \infty}}]{a.s.}
\int (t\log t - t \log s) \dd \mu_{\MP}(t) \dd \mu_{\MP}(s) = 
\frac32.
\end{split}
\label{kl32}
\end{equation}
In general the relative entropy is not symmetric, but for two generic states,
belonging to the same unitary orbit one has $S(\rho \| \sigma)=S(\sigma \|
\rho)$. In the case when one of the sates is maximally mixed  the symmetry is
broken and we get $ S(\rho \| \rho_*) \xrightarrow[{\tiny{N\to \infty}}]{a.s.}
\frac12$, while
$S(\rho_* \| \rho) \xrightarrow[{\tiny{N\to \infty}}]{a.s.} 1$. Results
(\ref{chernoff}) and (\ref{kl32}), compatible with \cite{Mo07}, combined with
the quantum Sanov \cite{BD+08} and Chernoff bounds \cite{Cas08,ANSV08} imply: %

\begin{proposition}.
Perform $m$ measurements on a generic quantum state $\rho$ of a large dimension.
Probability of obtaining results compatible with measurements performed on
another generic state $\sigma$ scales as $\exp(-3m/2)$. In a symmetric setup,
probability of erroneous discrimination between both states behaves as
$\exp(-64m/9\pi^2)$.
\end{proposition}

%%%%%%%%%%%%%%%%%%%%%%%%%%%%%%%%%%%%%%%%%%%%%%%%%%%%%%%%%%%%%%%%%%%%%%%%%%%%%%%%
\section{Transmission distance and Bures distance}
\label{TranBures}
%%%%%%%%%%%%%%%%%%%%%%%%%%%%%%%%%%%%%%%%%%%%%%%%%%%%%%%%%%%%%%%%%%%%%%%%%%%%%%%%
The Jensen--Shannon divergence (JSD) characterizes concavity of the entropy.
Its quantum analogue  (QJSD) deals with the von Neumann entropy, $S(\rho)=-{\rm
Tr}\rho \log \rho$, and is equal to the Holevo quantity $\chi$,
\begin{equation}
QJSD(\rho,\sigma) := S\left[ (\rho + \sigma)/2 \right] - 
[S(\rho) + S(\sigma)]/2 .
\end{equation}
Level density of the sum of two generic random states, $\rho+\sigma$, is
asymptotically described by the $\MP_2$ distribution. As the average entropy
for $\MP_c$ distribution for $c \geq 1$ reads \cite{page1993average},
\begin{equation}
\langle S \rangle_c = -\int x \log x \dd
\mu_{\MP_c(x)} = -\frac12 -  c \log c ,
\end{equation}
using Eq.~\eqref{integra} it is easy to
show  that
\begin{flalign}
&\mathrm{QJSD}(\rho,\sigma)  = 
-\frac12 \int t [\log t - \log 2N] \dd  \mu_N^{\rho+\sigma}(t)\\
& \nonumber
+ \frac12 \int t [\log t - \log N] 
[ \dd  \mu_N^{\rho}(t) + \dd \mu_N^{\sigma}(t)] 
\xrightarrow[{N\to \infty}]{a.s.}
\\
&\nonumber  
-\frac12 \int t \log t \dd  \mu_{\MP_2}(t) + \log 2
+ \int t \log t \dd  \mu_{\MP}(t) = \frac14.
\end{flalign}
JSD is a symmetric function of both arguments but it does not satisfy the
triangle inequality. However, its square root is known \cite{ES03} to yield a
metric for classical states, which leads to the transmission distance
\cite{BH09}, $D_T(\rho, \sigma) = \sqrt{\mathrm{QJSD}(\rho,\sigma)} \to 1/2$.

Squared overlap of two pure states, $|\langle\psi|\phi\rangle|^2$, can be
interpreted in terms of probability. A suitable generalization of this notion
for mixed states, called \emph{fidelity}, is defined in (\ref{fidel}).
The average fidelity and root fidelity 
between random states for small $N$ was analyzed in \cite{ZS05}. 
To cope with the same problem for large dimensions
we use the Fuss--Catalan distribution \cite{PZ11},
\begin{flalign}
&\mathcal{FC}(x) =  \frac{\sqrt[3]{2} \sqrt{3}}{12 \pi} \;
 \frac{\bigl[\sqrt[3]{2} \left(27 + 3\sqrt{81-12x} \right)^{\frac{2}{3}} -
   6\sqrt[3]{x}\bigr] } {x^{\frac{2}{3}}
     \left(27 + 3\sqrt{81-12x} \right)^{\frac{1}{3}}},\!\!\!\!\!\!\!\!&
\label{pi2}
\end{flalign}
which is supported in $[0,27/4]$ and describes the asymptotic 
level density of a product
$\rho\sigma$ of two random density matrices. Hence, the average root fidelity
can be asymptotically expressed as an integral over the Fuss-Catalan measure,
\begin{equation}
\sqrt{F(\rho, \sigma)} = \sum_i \sqrt{\lambda_i (\rho \sigma)}
\to \int \sqrt{x} \;
{\mathcal{FC}}(x)
\dd x = \frac34.
\label{rootfidel}
\end{equation}
This important result, compatible with~\cite{ZS05}, determines the limiting
behavior of the Bures distance \cite{BZ06},
\begin{flalign}
&\!\!\!\! D_B(\rho, \sigma) = \sqrt{2 (1 - \sqrt{F(\rho, \sigma)})} 
\to \sqrt{2 (1 - 3/4)} = \frac{\sqrt{2}}{2}.
\!\!\!\!\!\!\!\!\!\!\!\!\!\!\!\!\!
&
\end{flalign}
Observe that the known bounds \cite{ANSV08} between fidelity, Chernoff
information and trace distance, 
$F\le Q \le \sqrt{F} \le \sqrt{1-{\tilde D}^2}$, 
are generically satisfied with a healthy margin,
 $0.562 < 0.720 < 0.75 < 0.823$.

Making use of the Marchenko-Pastur distribution we 
calculate the average root  fidelity between a random state $\rho$ and the maximally mixed state $\rho_*$, 
\begin{equation}
\sqrt{F(\rho, \rho_*)} \to \int \sqrt{t} \dd \mu_\MP(t) = \frac{8}{3\pi}
\approx 0.8488 
\label{E35}
\end{equation}

In a similar way one can treat the related Hellinger distance \cite{BZ06},
$D_H(\rho,\sigma) = \sqrt{2 - 2  \tr \rho^{\frac12}\sigma^{\frac12}}$. Making
use of the average (\ref{chernoff}) we find the asymptotic behavior,
\begin{equation}
D_H(\rho,\sigma) \to \sqrt{2 - 2 \left(\frac{8}{3 \pi}\right)^2 } \simeq 0.7476 
\label{helling}
\end{equation}

\begin{figure}
\centering\includegraphics{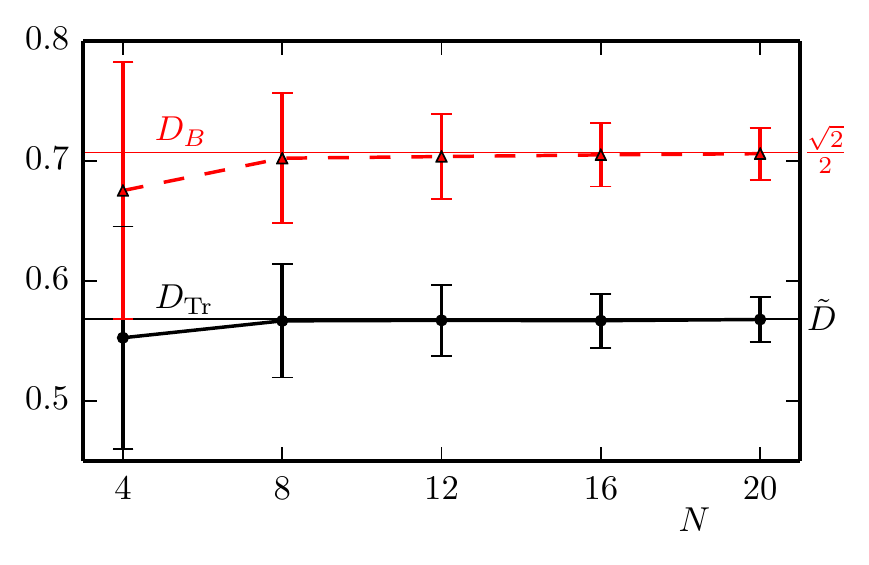}
\caption{Dependence of average distance between two generic states on the 
dimension $N$. Dashed (red) line shows the Bures distance and
 solid (black) line shows  the trace distance. The horizontal lines mark 
 the asymptotic values.}\label{fig:dim-dep}
\end{figure}

Dependences of averaged values of selected distances between two generic states
on the dimensionality shown in Fig.~\ref{fig:dim-dep} 
are consistent with rumors that
form the point of view of random matrix theory
 a dozen  is already close to infinity.

%%%%%%%%%%%%%%%%%%%%%%%%%%%%%%%%%%%%%%%%%%%%%%%%%%%%%%%%%%%%%%%%%%%%%%%%%%%%%%%%
\section{Average entanglement of quantum states}
%%%%%%%%%%%%%%%%%%%%%%%%%%%%%%%%%%%%%%%%%%%%%%%%%%%%%%%%%%%%%%%%%%%%%%%%%%%%%%%%
Generic pure quantum states are strongly entangled \cite{ZS01,HLW06}.
Integration over the $\MP$ measure may be applied to evaluate %the
average %degree of
entanglement of a random pure state $|\psi\rangle \in {\cal H}_N \otimes {\cal
H}_N$. Its  $G$--concurrence is defined \cite{Go05} as $G=N ({\rm
det}\rho)^{1/N}$, where $\rho= {\rm Tr}_N |\psi\rangle \langle \psi|$. For
large $N$ the concurrence converges almost surely to 
$\langle G\rangle  \to \exp [\int \log(x)
d \mu_{\MP}(x) ]=1/e$, in accordance with \cite{CSZ06}.

Results developed in the former sections allow us to study the overlap between
a given pure bipartite state $\ket{\psi} \in \Omega_N \otimes \Omega_N$ and the
nearest maximally entangled state. This overlap, also called maximal
entanglement fidelity, is related to another measure  $\mathcal{N}$ of
entanglement, called negativity. The relation is as follows
\begin{equation}
\begin{split}
\mathcal{N}(\proj{\psi}) &= \frac12 \left(
\left(\sum_{i} \sqrt{\lambda_i}\right)^2 -
1\right) =\\
&= \frac12 \left(
d \max_{\ket{\psi_+} \in \Omega^\mathrm{ME}_{N^2}} |\braket{\psi_+}{\psi}|^2 -
1\right),
\end{split}
\end{equation}
where $\Omega^\mathrm{ME}_{N^2}$ denotes the set of maximally entangled states
and $\lambda_i$ are the Schmidt coefficients of the state $\ket{\psi}$,
which satisfy $\sum_i \lambda_i =1$.
For a partial trace, $\rho = \tr_1 \proj{\psi}$,
we have
\begin{equation}
\sum_{i} \sqrt{\lambda_i} = \tr \sqrt{\rho},
\end{equation}
 Now we consider a random pure state distributed
according to the unitarily invariant Haar measure. 
Integral (\ref{E35})  % {rootfidel})
 implies that
\begin{equation}
\frac{1}{\sqrt{N}}\tr \sqrt{\rho} \to \frac{8}{3 \pi},
\end{equation}
almost surely when $N \to \infty$. 
The above result gives us the asymptotic behaviour
of the negativity for pure states
\begin{equation}
\mathcal{N}(\proj{\psi}) \simeq
\frac12 \left(
\left(\frac8{3\pi} \right)^2 N- 1\right).
\end{equation}
Formally we have an almost sure convergence in the asymptotic limit,  $N \to 
\infty$,
\begin{equation}
\frac{1}{N}\mathcal{N}(\proj{\psi}) \to \frac12 \left(\frac8{3\pi} \right)^2
\approx 0.3602
\end{equation}

Average entanglement of random  mixed states can also be analyzed
\cite{BTL12}. As shown by Aubrun \cite{aubrun2012partial} the level density of
partially transposed random states, written $\rho^{\mathrm{T}_A}$, of
a bipartite system obeys asymptotically the shifted semi-circular law of Wigner,
\begin{equation}
\lambda(\rho^{\mathrm{T}_A}) \sim \frac{1}{2 \pi c} \sqrt{4c - (x - c)^2}.
\end{equation}
Therefore, assuming $0<c<4$, the fraction $f_{\cal N}$ of negative eigenvalues
depends on the measure,
\begin{equation}
\begin{split}
 f_{\cal N} &=
\int_{c-2\sqrt{c}}^0 \frac{1}{2 \pi c} \sqrt{4c - (x - c)^2} \dd x \\
&=\frac{4\arccos\left( \frac{\sqrt{c}}{2}\right) - \sqrt{4c - c^2}}{4\pi},
\end{split}
\end{equation}
and tends to $0$ for $c\to 4$ in agreement with~\cite{aubrun2012partial}. 
Furthermore, the average negativity,
${\cal N}(\rho) = \tr |\rho^{{\mathrm{T}}_A} | - 1$, of a random
mixed state behaves as
\begin{equation}
\begin{split}
\cal N &=
\int_{c-2\sqrt{c}}^{0} 
\frac{|x|}{2\pi c}
\sqrt{4c - (x - c)^2} \dd x =\\
&=\frac{8\sqrt{4c - c^2} + \sqrt{4c^3-c^4}-12 c \arccos\left( 
\frac{\sqrt{c}}{2} 
\right)}{12 \pi}.
\end{split}
\end{equation}
For the flat measure, $c=1$, we get $f_{\cal N}=\frac13 - \frac{\sqrt{3}}{4\pi}
\simeq 0.1955$ and ${\cal N}=\frac{3\sqrt{3}}{4\pi} - \frac13 \simeq 0.080$
respectively, which implies that a generic bipartite state is weakly entangled.
The above numbers can be compared with the values for the maximally entangled
state, $\rho_+ = \frac1N  \sum_{ij} \ketbra{ii}{jj}$, which read $f_{\cal
N}(\rho_+) \to \frac12$ and ${\cal N}(\rho_+) = N-1.$

%%%%%%%%%%%%%%%%%%%%%%%%%%%%%%%%%%%%%%%%%%%%%%%%%%%%%%%%%%%%%%%%%%%%%%%%%%%%%%%%
\section{Classical probability vectors and average coherence}
%%%%%%%%%%%%%%%%%%%%%%%%%%%%%%%%%%%%%%%%%%%%%%%%%%%%%%%%%%%%%%%%%%%%%%%%%%%%%%%%
Let $q=(q_1, \dots q_N)$ be a normalized probability vector, so that
$q_i\ge 0$ and $\sum_i q_i=1$. A vector $q$ belongs to the probability simplex
of dimension $N-1$, in which one defines a family of \emph{symmetric
Dirichlet measures},
parameterized by a real index $s$, 
\begin{equation}
P_s(q) =  {\cal C}_s  \ {\delta}\left( \sum_{i = 1}^N q_i - 1\right)
  \; (q_1q_2 \dots q_N)^{s-1}.
\label{Dirichlet}
\end{equation}
Here ${\cal C}_s$ denotes a suitable normalization constant, which is dimension
dependent. Note that the flat measure in the simplex is recovered for $s=1$,
while the case $s=1/2$ corresponds to the statistical measure \cite{BZ06}.

Both of the above measures can be related to ensembles of random matrices. The
flat measure, $s=1$, describes distribution of squared absolute values of a
column (or a row) of a random matrix distributed according to the Haar measure
on the unitary group, while the statistical measure, $s=1/2$, corresponds to
random orthogonal matrices \cite{ZS01}. It is known~\cite{HZ90} that eigenstates
of matrices pertaining to circular unitary and circular orthogonal ensembles are
distributed according to the Haar measure on unitary and orthogonal groups,
respectively. Therefore, these distributions characterize statistics of
eigenvectors of random unitary matrices pertaining to circular unitary ensemble
(CUE) and  circular orthogonal ensemble (COE), respectively \cite{Me91}.

It is also possible to relate these measures with ensembles of random quantum
states. Writing a projector on a random pure state $|\psi\rangle$ in the
computational basis, (or the basis  state $|1\rangle \langle 1|$ in a random
basis) we arrive at a random projector matrix $\omega=|\psi\rangle \langle
\psi|=\omega^2$. The coarse graining channel describes decoherence, as it sends
any state $\rho$ into a diagonal matrix,
\begin{equation}
  \Phi_{CG}(\rho)=\sum_i |i\rangle  \langle i | \rho |i\rangle  \langle i |
    = {\rm diag}(\rho).
\label{coarse}
\end{equation}
Applying this channel to a projector $\omega$
we obtain a matrix with classical probability vector,
$q_i=\omega_{ii}=|\langle i|\psi\rangle |^2$,
at the diagonal.
If $|\psi\rangle$ represents a complex random vector,  
then the joint probability distribution $P(q)$ is given 
by the Dirichlet distribution (\ref{Dirichlet}) with $s=1$,
while random real vector leads to the case $s=1/2$.

Integrating out $N-1$ variables from the Dirichlet 
distribution (\ref{Dirichlet}) one obtains the distribution of a single
component $x=q_i$ of the probability vector.
For a fixed dimension $N$ these distributions read, 
for $s=1/2$ and $s=1$, respectively,
\begin{equation}\label{eqn:classical-measures}
\begin{split}
\tilde{P}_\mathrm{1/2}(x) & = \frac{\Gamma\left( \frac{N}{2} 
\right)}{\Gamma\left( 
\frac{N-1}{2} \right)} \frac{(1-x)^{(N-3)/2}}{\sqrt{\pi x}}, \\
\tilde{P}_\mathrm{1}(x) & = (N-1)(1-x)^{N-2}.
\end{split}
\end{equation}
In the asymptotic case, $N \to \infty$, the above distributions converge, after
an appropriate scaling, to the $\chi^2_{\nu}$ distribution with the number
$\nu$  of degrees of freedom equal to $1$ (orthogonal case) and $2$ (unitary
case)~\cite{HZ90}:
\begin{equation}
\begin{split}
P_\mathrm{1/2}(t) & =  \chi^2_1(t) = \frac{1}{\sqrt{2\pi t}} \ee^{-t/2},\\
P_\mathrm{1}(t) & =  2\chi^2_2(2t) = \ee^{-t}.
\end{split}
\label{eq:classical-dists}
\end{equation}
Here, $t= N x$, denotes a rescaled variable, such that $\langle t \rangle=1$, 
due to
the fact that an element $\langle x \rangle$ is proportional to $1/N$. The
former formula, also known as the Porter--Thomas distribution, describes statistics
of a squared real gaussian variable, while the latter one describes
distribution of the squared modulus of a complex gaussian variable.

Using the distributions~\eqref{eq:classical-dists} we calculate the asymptotic
mean $L_1$ distance between two probability vectors $p$ and $q$, distributed
according to the statistical $D^\mathrm{s}$, and the flat measure
$D^\mathrm{f}$, respectively. Furthermore, we find their root fidelity:
\begin{equation}
\sqrt{F(p, q)} = \sum_i \sqrt{p_i q_i}.
\end{equation}
This quantity is also known as the Bhattacharya coefficient \cite{BZ06}.
Finally, we compute the 
root fidelity and the Bures distance from the maximally mixed vector $p_*$.
We get the following results for the root fidelity $\sqrt{F^\mathrm{s}}$
and $\sqrt{F^\mathrm{f}}$ averaged over statistical and flat measures, 
respectively,
\begin{equation}
\begin{split}
\left\langle\sqrt{F^\mathrm{s}(p, p_*)} \right\rangle \to & \int \! \sqrt{t} 
P_\mathrm{1/2}(t) \dd t = \sqrt{\frac{2}{\pi}} \approx 0.7979,\\
\left\langle\sqrt{F^\mathrm{f}(p, p_*)}\right\rangle \to & \int \! \sqrt{t} 
P_\mathrm{1}(t) \dd t= \frac{\sqrt{\pi}}{2} \approx 0.8862,\\
\left\langle\sqrt{F^\mathrm{s}(p, q)}\right\rangle \to & \int \! \sqrt{ts}
P_\mathrm{1/2}(t)P_\mathrm{1/2}(s) \dd t \dd s = \frac{2}{\pi} \approx 0.6366,\\
\left\langle\sqrt{F^\mathrm{f}(p, q)}\right\rangle \to & \int \! \sqrt{ts}
P_\mathrm{1}(t)P_\mathrm{1}(s) \dd t \dd s = \frac{\pi}{4} \approx 0.7853.
\end{split}
\label{oeue}
\end{equation}
Observe that for both measures the average root fidelity between two random
probability vectors asymptotically equals the average fidelity with respect to 
the uniform vector,
$\langle \sqrt{ F(p, q)}\rangle = \langle \sqrt{F(p, p_*)}\rangle^2 = \langle  
F(p, p_*)\rangle$.

The above numbers can be directly compared with results obtained in
Sec.~\ref{TranBures} for the quantum case. For instance, the average root
fidelity between two random classical states distributed according to the flat
measure, $(0.78)$, is larger than the analogous result, $(0.75)$, for quantum
states. Hence the average Bures distance satisfies the reversed inequality,
$\langle D_B(p,q)\rangle  <  \langle D_B (\rho, \sigma)\rangle$.

Mean $L_1$ distances between classical states averaged over the probability
simplex can be  expressed as integrals over the
measures~(\ref{eqn:classical-measures}). For large $N$ one can use asymptotic
measures~(\ref{eq:classical-dists}), which yield the following results,
\begin{equation}
\begin{split}
\frac12 \left\langle D^\mathrm{s}(p, p_*) \right\rangle & \to 
\frac12 \int |t - 1| P_\mathrm{1/2}(t) \dd t =
 \sqrt{\frac{2}{\pi \ee}}, \\
\frac12 \left\langle D^\mathrm{f}(p, p_*) \right\rangle & \to 
\frac12 \int |t - 1| P_\mathrm{1}(t) \dd t =
\frac{1}{\ee},\\
\frac12 \left\langle D^\mathrm{s}(p, q) \right\rangle & \to 
\frac12 \int |t-s| P_\mathrm{1/2}(t)P_{1/2}(s) \dd t
\dd s = \frac{2}{\pi},\\
\frac12 \left\langle D^\mathrm{f}(p, q) \right\rangle & \to
\frac12 \int |t-s| P_\mathrm{1}(t)P_{1}(s) \dd t
\dd s = \frac12 .
\end{split}
\end{equation}
To make a direct comparison with the results obtained in the quantum case
easier, each formula above contains the same prefactor $1/2$, present in the
definition (\ref{traced}) of the trace distance.

In order to quantify coherence of a state $\rho$ with respect to a given basis
one uses the \emph{relative entropy of coherence} \cite{BCP14}, defined as the
difference between the entropy of the diagonal of the density matrix and its
von Neumann entropy,
\begin{equation}
C_{\rm rel.ent} (\rho) =  S\bigl(\Phi_{CG}(\rho)\bigr) - S(\rho).
\label{crelent}
\end{equation}
It belongs to $[0,\log N]$ and is equal to the increase of the entropy of a
state $\rho$ under the action of the coarse graining channel (\ref{coarse}).

The maximal value, $C_{\rm rel.ent} = \log N$, is achieved for contradiagonal 
states~\cite{LPZ14}, for which $\rho_{ii} = 1/N$. If $U_{\min}$ denotes the 
unitary matrix of eigenvectors of $\rho$, so that $\rho_D = U_{\min}^\dagger 
\rho U_{\min}$ is diagonal, then $\rho_C = U_{\max}^{\dagger} \rho  U_{\max}$
is contradiagonal in the related basis $U_{\max} = U_{\min} F$ for any unitary 
complex Hadamard matrix $F$, which satisfies 
$F F^{\dagger} = \1$ and $|F_{ij}| = 1/N$~\cite{TZ06}.

For any pure state the second term in (\ref{crelent})
 vanishes, so for a complex random pure state
of a large dimension $N$ its average entropy of coherence equals
\begin{equation} \label{eqn:rel-entr}
\begin{split}
\langle &C_{\rm rel.ent} (|\psi_{\C}\rangle ) \rangle  =  
\langle  S (p) \rangle_{CUE} =\\
& = 
  - \int t \log t P_\mathrm{1}(t) \dd t = \log N - (1-\gamma),
\end{split}
\end{equation}
where $\gamma\approx 0.5772$ denotes the Euler constant. An analogous result
for a random real pure state reads
\begin{equation} \label{creorto}
\begin{split}
\langle &C_{\rm rel.ent} (|\psi_{\R}\rangle ) \rangle  =  
\langle  S (p) \rangle_{COE} =\\
&= 
  - \int t \log t P_\mathrm{1/2}(t) \dd t =  \log N - (2-\gamma - \log 2),
\end{split}
\end{equation}
where the average entropies were studied first by Jones~\cite{Jo90}. Thus a
random state is characterized with a high degree of coherence, close to the
maximal one, $\log N$. Interestingly, this statement holds almost surely for
any choice of the basis, with respect to which the decoherence takes place. We 
also note that these results are compatible with the ones presented 
in~\cite{SZP15} when we let $N \to \infty$.

For comparison let us consider a random mixed state $\rho$ distributed
according the Hilbert--Schmidt measure. Its average entropy is close to the
maximal, as it reads \cite{BZ06} for large dimensions, $\langle
S(\rho)\rangle_{HS} =\log N -1/2$. Therefore, its relative entropy of
coherence, equal to the entropy gain induced by the decoherence channel
(\ref{coarse}) cannot be large. In the generic bases the diagonal elements of a 
complex Wishart
matrix are asymptotically distributed according to the $\chi^2_{\nu}$
distribution with $\nu=2N$ degrees of freedom, the entropy of the diagonal
behaves as $ S ({\rm diag}(\rho)) \simeq \log N - 1/2N$. Therefore, the
relative entropy of coherence of a random mixed state $\rho$ asymptotically
tends to a constant,
\begin{equation}
\langle C_{\rm rel.ent} (\rho) \rangle  =  
\langle  S ({\rm diag}(\rho))  \rangle -
\langle S(\rho)\rangle_{HS} \to 1/2,
\label{cremix}
\end{equation}
characteristic to a random mixed states in a contradiagonal form, 
for which  $C_{\rm rel.ent}$ is maximal.
Thus the average coherence of random states decreases with their purity, as
expected: It behaves asymptotically as $\log N$ for a random pure state and it
is equal to a constant for a random mixed state. This result is compatible with 
the one presented in Tab. II in~\cite{ZSP15}, when we multiply the values in 
the second column by $\log N$.

Note that result \eqref{eqn:rel-entr} can be used to explain recent findings
of~\cite{ISPU15}. The mean entropy $\langle S_{\rm diag}(\tau) \rangle$ of the
diagonal of a quantum state averaged over a sufficiently long time $\tau$ tends
to the average over the ensemble of random states, which in the case of complex
pure states tends to the mean entropy~\eqref{eqn:rel-entr} over the flat measure
on the simplex. It is equal to $\log N-(1-\gamma)$ and becomes larger for mixed
states. As the entropy of the time averaged state $S_{\bar \rho}$ is limited by
$\log N$, their difference $\Delta S= S_{\bar \rho} - \langle{S_{\rm
diag}(\tau)}\rangle$ for a generic unitary evolution satisfies the bound $\Delta
S \le 1-\gamma$, in agreement with \cite{GGM16}. However, taking an initial real
pure state $\rho_R$ and restricting attention to purely imaginary generators,
$H=i A$ with $A$ real antisymmetric, $A=-A^T$, one obtains an orthogonal evolution
matrix, $O=e^{iH}=e^{-A}$, such that the state $\rho_R'=O\rho_R O^T$ remains real.
In such a case the average entropy of the diagonal is equal to the entropy
averaged over the statistical measure~\eqref{creorto}, so that the bound for the
entropy difference becomes weaker,
\begin{equation}
\Delta S \le 2- \gamma- \log 2 \approx 0.7297
\end{equation}

Next we wish to calculate the $L_1$ coherence of a generic quantum state
\begin{equation}
C_{L_1} = \sum_{i \neq j} |\rho_{i,j}|.
\end{equation}
In order to achieve this observe first that for any $N > 1$
and a random pure state
$\rho=\proj{\phi}$ distributed according to the Haar measure, we have:
\begin{equation}
|\rho_{ij}| = \sqrt{|\phi_i \phi_j|^2}.
\end{equation}
Next, we note that the vector $q=[|\phi_1|^2,\ldots, |\phi_N|^2]$ pertains the
$N$-dimensional symmetric Dirichlet distribution. For real states we get $q
\sim P_{1/2}(q)$ and for complex states we get $q \sim P_1(q)$. Next, we note
that $\mathbb{E}(|\rho_{ij})| = \mathbb{E}(\sqrt{q_i}\sqrt{q_j})$. To calculate
the above expectational value we will use a formula for mixed moments of the
Dirichlet distribution. For a random vector $q$ distributed according to the
Dirichlet distribution with parameters given by the vector $\alpha$ we
have~\cite{NTT11}
\begin{equation}
\mathbb{E}\left( \prod_{i=1}^N q_i^{\beta_i} \right) = 
\frac{\Gamma(\sum_{i=1}^{N} \alpha_i )}{\Gamma(\sum_{i=1}^{N} 
\alpha_i+\beta_i)} \times \prod_{i=1}^N \frac{\Gamma(\alpha_i + 
\beta_i)}{\Gamma(\alpha_i)}.
\end{equation}
To calculate the expected coherence of a pure state we set $\alpha_i = 1$ for 
complex states and $\alpha_i=1/2$ for real vectors. In both cases we set 
$\beta_1=\beta_2=1/2$ and $\beta_i=0$ for $i \geq 3$. We also note that for 
large $N$, we have $C_{L_1} \simeq N(N-1) \langle |\rho_{ij}| \rangle$. We 
obtain
\begin{equation} \label{eqn:CL1-pure}
\begin{split}
C_{L_1}^\R (\psi_{\R})& = (N-1) \frac{2}{\pi},\\
C_{L_1}^\C (\psi_{\C})& = (N-1) \frac{\pi}{4}.
\end{split}
\end{equation}
Values obtained above are related to the limiting values of the root fidelity
for classical probability vectors presented in Eqn.~\eqref{oeue}. This happens
because in the asymptotic scenario the distribution of the probability
components and non-diagonal elements of a pure state, differ only by a scaling
factor.

These values can be compared with the maximal value attained for pure state in 
a contradiagonal form, $\rho^{\psi}_{\C}$ for which $|(\rho^{\psi}_{\C})_{ij}| 
= 1/N$, so that $C_1(\rho^{\psi}_{\C}) = N(N-1)\frac1N = N-1$. This result is 
greater, than the mean value~\eqref{eqn:CL1-pure} for a complex random state in 
a generic basis, by a factor $\pi /4 \sim 0.785$.

\begin{figure}
\subfloat{\centering\includegraphics[width=0.23\textwidth]{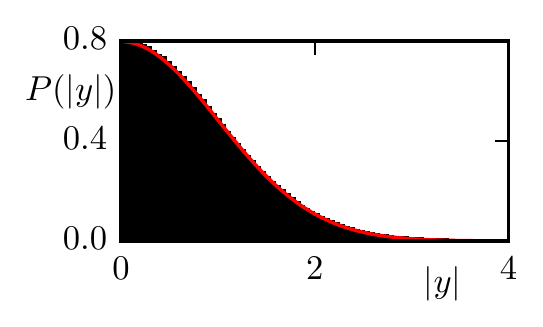}}
\subfloat{\centering\includegraphics[width=0.23\textwidth]{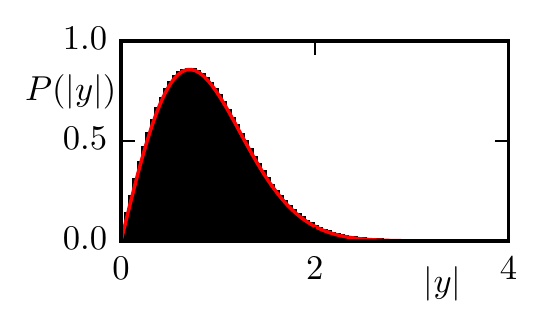}}
\caption{Histogram of absolute values of rescaled off-diagonal elements $y = 
\rho_{ij} $ of random mixed state. Left real state and right complex state. The 
solid (red)
 curves represent the distributions $\chi_1(|y|)$ (left) and $\sqrt{2} 
\chi_2(\sqrt{2}|y|)$ (right).} 
\label{fig:mixed-coherence}
\end{figure}

The random mixed state $\rho= G G^{\dagger}/ \tr G G^{\dagger}$ can be
considered as a normalized Gram matrix for an ensemble of $N$ random complex
vectors. In the asymptotic limit the normalization pre-factor $N /\tr G
G^{\dagger}$ tends to a constant. Thus in the limiting case we may consider
\begin{equation}
\rho_{ij} = \frac1N \sum_{k=1}^N G_{ik} \overline{G}_{jk}.
\end{equation}
From the central limit theorem we get that $\rho_{ij}$ has a normal distribution
for $N \gg 1$. This implies that the absolute value 
$|\rho_{ij}|$  is described, up to a scaling factor,
by an appropriate $\chi_{\nu}$ distribution. It is the distribution of the square
root of the sum of squares of independent random variables having a standard
normal distribution~\cite{EHP2000}. The $\chi_{\nu}$ 
distribution has a single parameter,
the number $\nu$ of degrees of freedom, equal to the number of summed normal
variables. Hence, we get:
\begin{equation}
\begin{split}
P_\R(y) &= \chi_1(y) = \sqrt{\frac{2}{\pi}}\ee^{-y^2/2} \\
P_\C(y) &= \sqrt{2} \chi_2(\sqrt{2}y) = 2 x \ee^{-y^2},
\end{split}
\end{equation}
where $y = \sqrt{N} |\rho_{ij}|$. 
Fig.~\ref{fig:mixed-coherence} shows that statistical distribution of modulus
of off-diagonal elements of a random mixed state which can be asymptotically 
described by the $\chi$ distribution.
This property allows us to describe asymptotic behavior of the coherence of 
real and complex random mixed states,
\begin{eqnarray}
C_{L_1}^\R(\rho^{\R}) & \simeq& \sqrt{N} \int_0^\infty x P_\R \dd x  = 
\sqrt{N}  \sqrt{\frac{2}{\pi}}, \label{eqn:CL1-mixed-R}\\
C_{L_1}^\C(\rho^{\C}) & \simeq& \sqrt{N} \int_0^\infty x P_\C \dd x = \sqrt{N}  
\frac{\sqrt{\pi}}{2}. \label{eqn:CL1-mixed-C}
\end{eqnarray}

In the above, we have shown that the $L_1$ coherence, $C_{L_1}$, scales as
the dimension  $N$  for generic pure states
 and as $\sqrt{N}$ for typical mixed states.
Note that expression~\eqref{eqn:CL1-mixed-C} asymptotically holds also for 
contradiagonal states, for which $\rho_{ii} = 1/N$.

%%%%%%%%%%%%%%%%%%%%%%%%%%%%%%%%%%%%%%%%%%%%%%%%%%%%%%%%%%%%%%%%%%%%%%%%%%%%%%%%
\section{Dynamical system--coupled quantum kicked tops}
%%%%%%%%%%%%%%%%%%%%%%%%%%%%%%%%%%%%%%%%%%%%%%%%%%%%%%%%%%%%%%%%%%%%%%%%%%%%%%%%
To show a direct link with the theory of quantized chaotic systems we analyze
the model of quantum kicked top. It is described by the angular momentum
operators $J_x, J_y, J_z$, satisfying standard commutation rules. The dynamics
of a single system consists of a periodic unitary evolution in the Hilbert
space of dimension $N=2j+1$, followed by an infinitesimal perturbation
characterized by the kicking strength $k$. To analyze the effects of quantum
entanglement the model consisting of two coupled tops described by the
Hamiltonian, $H(t) = H_1(t) + H_2(t) + H_{12}(t)$, with \; $H_i(t) =  p_i
J_{y_i} + \frac{k}{2j_i} J_{z_i}^2 \sum_n \delta(t-n)$, $H_{12}(t) = \epsilon /
{\bar j} \left( J_{z_1} \otimes J_{z_2} \right) \sum_n \delta(t-n)$ and ${\bar
j}=\frac{j_1+j_2}{2}$. This model was studied in \cite{MS98,BL04,DDK04}.

We set the standard values of the parameters $p_1=p_2 = \pi /2$  and consider
the unitary one--step time evolution operator, $U =  U_{12} (U_1 \otimes U_2)$,
where
\begin{equation}
\begin{split}
 U_i = & \exp\left( -\ii \frac{k}{2j_i}J_{z_i}^2 \right)
\exp \left( -\ii \frac{\pi}{2} J_{y_i} \right), \ \ i=1,2, \\
U_{12} = & \exp \left( -\ii \frac{\epsilon}{\bar j} J_{z_1} \otimes J_{z_2} \right).
\end{split}
\label{2tops}
\end{equation}
\begin{figure}[!h]
\subfloat{\centering\includegraphics[width=0.23\textwidth]{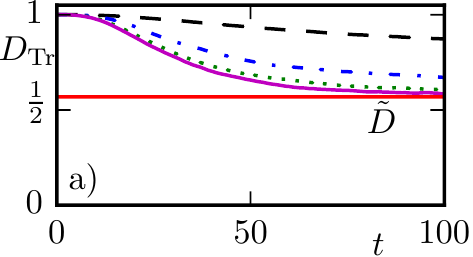}}
\subfloat{\centering\includegraphics[width=0.23\textwidth]{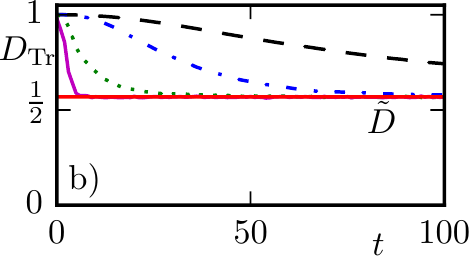}}
\caption{Trace distance $D_{\rm Tr}$ between initially orthogonal reduced states of the
kicked tops as a function of time $t$ obtained for $j_1 = j_2 = 60$. Convergence to
the asymptotic  value $\tilde D$ represented by the horizontal line depends on
a) kicking strength,  $k=3.2$ (upper dashed line), $k=3.5, 3.8$ and $k=4.0$
(solid line) and b) the coupling constant, $\epsilon=0.005$ (upper dashed
line), $\epsilon=0.01, \ 0.1$ and $\epsilon=1$ (solid line).}
\label{fig:chaos}
\end{figure}
Consider a pure separable state $|l\rangle \otimes |l\rangle$ evolved unitarily
for $t$ steps, which gives $|\psi_l^t\rangle = U^t (|l\rangle \otimes 
|l\rangle)$ for $l=1,2$. After the unitary dynamics we perform partial trace 
over the second subsystem, thus defining two states:
\begin{equation}\label{eqn:traced-tops}
\sigma_1(t) = \tr_2 \proj{\psi^t_1} 
{\ \ \ \rm and \ \ \ }
\sigma_2(t) = \tr_2 \proj{\psi^t_2},
\end{equation}
and study the spectrum of their difference $\Gamma = \sigma_1 -\sigma_2$. Level
density of $\Gamma$  obtained for $k=6$, 
corresponding to the chaotic regime, with Wigner level spacing statistics,
$\epsilon=0.01$ and four values of the
 dimensionality ratio, $c=(2j_2+1)/(2j_1+1)$, can be described by distribution
(\ref{eq:sym-mp-dist}) as shown in Fig.~\ref{fig1}. This observation confirms
the conjecture that a quantized deterministic chaotic system may lead to
generic, random states.
Consider the trace distance $D_{\rm Tr} \bigl( \sigma _1(t), \sigma _2(t)
\bigr)$, which  at $t=0$ is equal to unity as the initial states are
orthogonal. For larger times $t$ the  trace 
distance tends to the % universal
number $\tilde D$ derived in (\ref{aveboth}) and the convergence rate is faster
for strongly chaotic systems (large $k$) and large coupling strength
$\epsilon$ -- see Fig. \ref{fig:chaos}.

The trace distance $D_{\rm Tr} \bigl( \sigma _1(t), \sigma _2(t) \bigr)$
between two initially orthogonal states undergoing the dynamics of the coupled
kicked top (\ref{2tops}) decreases monotonously and tends to the universal
number $\tilde D$. The rate of convergence to the asymptotic value depends on
the parameters of the model, as shown in Fig. \ref{fig:chaos}. However, if we
repeat the procedure taking for initial states the tensor products of two
coherent states localized nearby, the trace distance between reduced states
initially grows in time. This feature is related to the fact that the dynamics
taking place in the reduced system is not Markovian \cite{LPB10} and that there
exist correlations between both systems.

Note that the investigated state of the first subsystem is obtained by a
reduction over the second subsystem, $\sigma (t)={\rm Tr}_2|\psi(t)\rangle
\langle \psi(t)|$. Thus to obtain the next iteration $\sigma(t+1)$, one needs
to purify $\sigma(t)$ in a specific way to get the bipartite pure state
$|\psi(t)\rangle$, evolve it unitarily into $|\psi(t+1)=U|\psi\rangle$ and
perform the partial  trace, $\sigma (t+1)={\rm Tr}_2|\psi(t+1)\rangle \langle
\psi(t+1)|$. The purification procedure is not unique, and the information
encoded in $\sigma$ is not sufficient to recover the desired pure state
$|\psi(t)\rangle$. Hence the dynamics of the reduced state, $\sigma(t) \to
\sigma(t+1)$, is not Markovian. Furthermore, the quantum dynamics is explicitly
defined for a full system consisting of two coupled tops, but the reduced
dynamics of a single subsystem is not well defined independently of the
correlations.

%%%%%%%%%%%%%%%%%%%%%%%%%%%%%%%%%%%%%%%%%%%%%%%%%%%%%%%%%%%%%%%%%%%%%%%%%%%%%%%%
\section{Concluding remarks}
%%%%%%%%%%%%%%%%%%%%%%%%%%%%%%%%%%%%%%%%%%%%%%%%%%%%%%%%%%%%%%%%%%%%%%%%%%%%%%%%
We evaluated common distances between two generic quantum states. Due to the
concentration of measure in high dimensions these distances converge to
deterministic values -- see Tab.~\ref{tab:summary}. Our results are directly
applicable for quantum hypothesis testing \cite{ANSV08} as they imply concrete
bounds on the distinguishability between generic states. Asymptotic value of the
root fidelity, $\sqrt{F}=3/4$, can be used as a universal benchmark for future
theoretical and experimental studies based on this quantity and Bures distance.

Although obtained expectation values are exact in the asymptotic limit our
numerical results presented in Fig.~\ref{fig:dim-dep} show that they can be used
for $N$ of the order of ten. Furthermore, results obtained improve our intuition
concerning the structure of the body $\Omega_N$ of quantum states and describe
coherence with respect to a generic basis of a typical pure and mixed quantum
state. Moreover, we demonstrated that a typical mixed state of a large
bipartite system is weakly entangled.

Results presented in this paper are obtained with use of the free probability
calculus and hold in the asymptotic limit $N\to \infty$. It would be therefore
interesting to apply other methods to obtain finite size corrections to the
expressions derived in this work and to show analytically, for what system
sizes the correction terms can be neglected in practice. In the case of the
$\MP$  distribution this has been studied in~\cite{F12}, although these results
do not directly apply in the case studied in this work. The free convolution
works only in the asymptotic limit, thus obtaining a distribution similar to
$\SMP$ for finite $N$ requires careful treatment. The problem of estimating the
average coherence for quantum systems of a finite size was recently studied 
in~\cite{ZSP15,SZP15} and their results are consistent with our asymptotic
expressions.

\begin{table}[!h]
\begin{tabular}{|c|c|c|c|}
\hline
x & $D_{\rm x}(\rho, \frac1N \1)$ & $D_{\rm x}(\rho, \sigma)$ 
& $D_{\rm x}(\ket{\psi},
\ket{\phi})$  \\ \hline
$\tr$ & $\frac{3\sqrt{3}}{4\pi} \approx 0.414$ & $\frac14 + \frac{1}{\pi} 
\approx 0.568$ & 1 \\ \hline
HS & 0 & 0 & 1\\ \hline
$\infty$ & 0 & 0 & 1 \\ \hline
T & $T_1 \approx 0.368$ & $\frac12$ & $\sqrt{\ln 2} \approx 0.833$ \\ \hline
B & $\sqrt{2 - \frac{16}{3\pi}} \approx 0.550$ & $\frac{\sqrt{2}}{2} \approx 
0.707$ & $\sqrt{2} \approx 1.414$ \\ \hline
E & $E_1 \approx 0.518$ & $\frac{1}{2\sqrt{2}} \sqrt{\ln\frac{8^8}{7^7}}
\approx 0.614$ & $\sqrt{\ln 2} \approx 0.833$ \\ \hline
H & $\sqrt{2 - \frac{16}{3\pi}} \approx 0.550$ & $\sqrt{2 - 2 \left(
\frac{8}{3\pi} \right)^2} \approx 0.748$ & $\sqrt{2} \approx 1.414$  \\ \hline
\end{tabular}
\caption{Typical distances $D_{\rm x}$ between generic mixed states $\rho$ and $\sigma$ and
pure states $|\psi\rangle$ and $|\phi\rangle$ of a large dimension $N$. 
Definitions of $D_{\tr}, D_T, D_B$ and $D_H$ are given in the text, while 
%We use the Hilbert--Schmidt distance 
$D_{\rm HS}(\rho, \sigma) = [\frac{1}{2} \tr(\rho - \sigma)^2]^{1/2}$,
and $D_\infty(\rho, \sigma) = \max_i |\lambda_i(\rho -
\sigma)|$, and the entropic distance \cite{LPS09} is $D^2_E(\rho, \sigma) =
H_2\left(\frac12 (1 - \sqrt{F(\rho, \sigma)})\right)$, with $H_2(x)
= -x \ln x - (1 - x)\ln(1 - x)$. Analytical formulas for the numbers $T_1$
and $E_1$ are provided in Appendix~\ref{supp:sec:entropic}.}
\label{tab:summary}
\end{table}

\section*{Acknowledgements}
It is our pleasure to thank Marek Ku{\'s} for more than 25 years of
enlightening discussions on quantum chaos, random matrices and quantum
entanglement. We are also grateful to G.~Aubrun, M.~Bo{\.z}ejko, A.~Buchleitner,
M.~Horodecki, P.~Horodecki, A.~Lakshminarayan, A.~Montanaro, I.~Nechita,
 M.~A.~Nowak,  P.~{\'S}niady, R.~Speicher, and A.~Szko{\l}a for helpful remarks and fruitful
correspondence. Further discussions  with SMP students on SMP distributions are
appreciated. This work was supported by the Polish National Science center
under projects number DEC-2012/05/N/ST7/01105 ({\L}P), DEC-2012/04/S/ST6/00400
(ZP) and DEC-2011/02/A/ST1/00119 (K{\.Z}).

\appendix
\section{Limiting behavior of functions of two independent states}
\label{supp:sec:functions}

In this section we consider the limiting value of the trace of a product of two
functions on random states. Let $\rho$ and $\sigma$ be random mixed states
generated according to the flat measure $\nu_{\rm HS}$ in $\Omega_N$ with
eigenvalues described by the measures $\mu_N^{(\rho)} = \frac1N \sum_i
\delta_{N \lambda_i(\rho)}$ and $\nu_N^{(\rho)} = \frac1N \sum_i \delta_{N
	\lambda_i(\sigma)}$. 
For large $N$ with probability one the measures $\mu_N$
and $\nu_N$ converge weakly to $\mu_{\MP}$. Because $\rho$ and $\sigma$ are
asymptotically free random matrices, thus for any analytic functions $g$ and
$h$ the matrices $g(\rho)$ and $h(\sigma)$ are also asymptotically free. The
product $g(\rho) h(\sigma)$ has a limiting, non-random distribution of
eigenvalues, given by an appropriate multiplicative free convolution. As a
result the normalized trace of the product tends, almost surely, to a
deterministic quantity, equal to the mean
\begin{equation}
\tr g(\rho) h(\sigma) 
 \xrightarrow[{N\to \infty}]{a.s.} 
 m,
\end{equation}
where
$
m = \lim_{N \to \infty }\mathbb{E}\tr g(\rho) h(\sigma).
$

Now we will calculate the limiting value of the mean trace of the product.
The eigendecomposition gives us 
\begin{equation}
\mathbb{E} \tr g(\rho) h(\sigma) =
\mathbb{E} \sum_{ij} g\left[\lambda_i(\rho)\right] 
h\left[\lambda_j(\sigma)\right] |U_{ij}|^2,
\end{equation}
where $U$ is a unitary transition matrix from eigenbasis of $\rho$ to 
eigenbasis of $\sigma$. The independence of eigenvectors distribution from
eigenvalues distribution gives us 
\begin{equation}
\mathbb{E} \tr g(\rho) h(\sigma) =
\mathbb{E} \sum_{ij} g\left[\lambda_i(\rho)\right] 
h\left[\lambda_j(\sigma)\right] \left(\mathbb{E} |U_{ij}|^2\right).
\end{equation}
Since the distribution of $\rho$ and $\sigma$ is invariant with respect to
unitary rotations, $U$ has a Haar distribution, for which we have, by the
permutation symmetry, $\mathbb{E} |U_{ij}|^2 = \frac1N$. This gives us
\begin{equation}
\mathbb{E} \tr g(\rho) h(\sigma) =
\frac1N  \mathbb{E} \tr g(\rho) \mathbb{E} \tr h(\sigma).
\end{equation}
Now we can make use of the weak convergence of the eigenvalues distributions 
similarly to~\eqref{integra} and get
\begin{equation}
\mathbb{E} \tr g(\rho) h(\sigma) 
 \xrightarrow[{N\to \infty}]{a.s.}
\int B_{g,h}(s,t)\dd \mu_{\MP}(t) \mu_{\MP}(s),
\end{equation}
where
\begin{equation}
B_{g,h}(s,t) = \lim_{N \to \infty} N g\left(\frac{s}{N}\right) 
h\left(\frac{s}{N}\right).
\end{equation}

\section{Distances as a function of the rectangularity parameter}
\label{appenB}

In this section we present the average trace and the rescaled Hilbert-Schmidt
distance between random states distributed with respect to the induced measure
$\nu_K$. It is convenient to parameterize the measure by the rectangularity
parameter $c=K/N$, where $N$ stands for the dimension of the principal system,
while $K$ denotes the dimension of the environment.

\begin{figure}[!h]
\subfloat[$D_\mathrm{Tr}$]{\centering\includegraphics{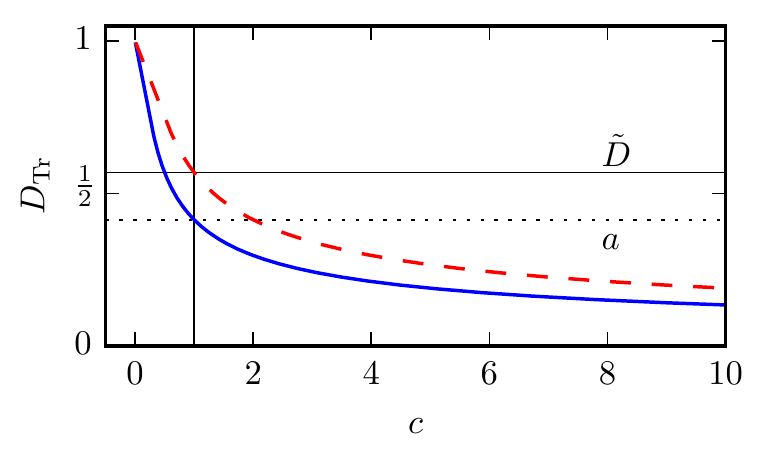}\label{fig:aa}}\\
\subfloat[$\sqrt{N} 
D_\mathrm{HS}$]{\centering\includegraphics{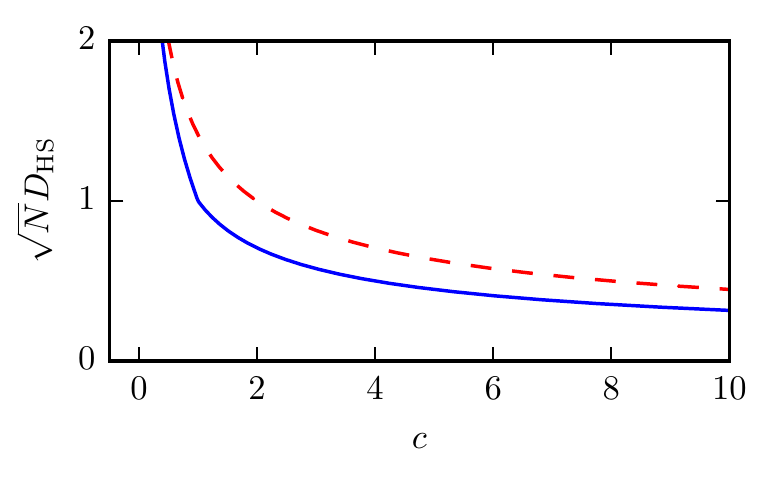}\label{fig:ab}}
\caption{Trace distance \protect\subref{fig:aa} and rescaled 
Hilbert-Schmidt distance \protect\subref{fig:ab} as a function of the
rectangularity parameter $c$. Solid blue curve denotes the distance of a random
state $\rho$ from the maximally mixed state $\rho_*$. Dashed red curve shows
the distance between two random states. The vertical lines marks $c=1$. The
solid horizontal line marks $\tilde{D}$, the distance between two generic
states for $c=1$. The dashed horizontal line marks $a$, the distance of a
generic state form the maximally mixed state $\rho_*$.}\label{fig:rectan}
\end{figure}

Following~\eqref{integra} we integrate the function $|t-c|$ over the
Marchenko--Pastur distribution ${\cal MP}_c$ to get the average trace distance
$D_{\rm Tr}(\rho,\rho_*)$ between a random state $\rho$ and the center of the
set, $\rho_*={\mathbbm 1}/N$. In analogy to Eq.~\eqref{aveboth} computing the
average $|x|$ with respect to the symmetrized MP distribution ${\cal SMP}_c$
one obtains the mean value of the trace distance $D_{\rm Tr}(\rho,\sigma)$
between two random states.

These quantities for the trace distance are shown in in Fig.~\ref{fig:aa} as a
function of the rectangularity parameter $c$. In the limit $c\to \infty$ the
distribution ${\cal SMP}_c$ tends to the circular law rescaled by $1/\sqrt{2c}$
so the distance $D_{\rm Tr}(\rho,\sigma)$ tends to zero as
$4\sqrt{2}/(3\pi\sqrt{c})$. Note that above average improves the bound recently
derived \cite{BH15} in order to show that exponential decay of correlations
implies area law. Integrating the function $|t-c|$ over the $\mathcal{MP}_c$
distribution we obtain that the average trace distance from the maximally mixed
state $\rho_*$, $D_\mathrm{Tr}(\rho, \rho_*)$ tends to zero as $4/(3 \pi
\sqrt{c})$.

In order to calculate the rescaled Hilbert-Schmidt distance between two generic
states, $\sqrt{N} D_\mathrm{HS}(\rho, \sigma)$, we evaluate 
the second moment of the $\mathcal{SMP}_c$ distribution. 
In the limit $c\to\infty$ the distance
tends to zero as $\sqrt{2}/\sqrt{c}$. 
To obtain average % rescaled 
Hilbert-Schmidt distance of a generic state from the maximally mixed state, $\sqrt{N} D_\mathrm{HS}(\rho, \rho_*)$,
we integrate the function $(t-c)^2$ over the
Marchenko-Pastur distribution $\mathcal{MP}_c$.
 In the limit $c\to\infty$ this distance behaves as $1/\sqrt{c}$ 
-- see Fig.~\ref{fig:ab}. Note that for both distances 
in the limit $c\to\infty$ we get $D_x(\rho, \sigma)=\sqrt{2} D_x(\rho, \rho_*)$, where $x$ stands either for the
trace or the Hilbert-Schmidt distance.

\section{Transmission distance and entropic distance}
\label{supp:sec:entropic}

We provide here an analytical expression for two distances listed in
Tab.~\ref{tab:summary}. Square root of quantum Jensen-Shannon divergence (QJSD)
leads to the transmission distance \cite{BH09}, $D_T(\rho, \sigma) =
\sqrt{\mathrm{QJSD}(\rho,\sigma)}$. Taking one random state $\rho$ and the
maximally mixed state $\rho_*$ we find that for large dimension $N$ this
distance converges to
\begin{equation}
\begin{split}
T_1^2 =& [D_T (\rho,\rho_*)]^2 \to 
 \frac18+ \frac{\sqrt{5}}{16}+ \frac{15}{16}\log 2+ \\
&+\log \sqrt[16]{4870847-2178309 \sqrt{5}},
\label{eq:t1}
\end{split}
\end{equation}
so that $T_1\simeq 0.368$.
In a similar way we find the entropic distance 
 defined in the caption to the Table I,
\begin{eqnarray}
E_1 &=& D_E (\rho,\rho_*) \to \\
&\to& \sqrt{\frac{3 \pi  \log \left(\frac{36 \pi ^2}{9 \pi
^2-64}\right)-16 \coth ^{-1}\left(\frac{3 \pi }{8}\right)}{6 \pi 
}} = 0.518. \nonumber 
\label{eq:e1}
\end{eqnarray}

\bibliographystyle{apsrev}

\end{document}